\documentclass{sig-alternate-05-2015}

\begin{document}

\setcopyright{acmcopyright}



\conferenceinfo{CIKM '16}{Oct 24--28, 2016, Indianapolis, USA}

\acmPrice{\$15.00}

%
\conferenceinfo{CIKM}{'16 Indianapolis, USA}

\title{Ensemble Methods for Personalized E-Commerce Search Challenge at CIKM Cup 2016 }

\author{
%
%
{Chen Wu$^{1}$, Ming Yan$^{1}$, Luo Si$^{1}$}\\
      \affaddr{$^{1}$Alibaba Group Inc, 969 West Wenyi Road, Hangzhou 311121, China}\\
       \email{\{wuchen.wc, ym119608, luo.si\}@alibaba-inc.com}\\
}

\maketitle
\begin{abstract}
Personalized search has been a hot research topic for many years and has been widely used in e-commerce. This paper describes our solution to tackle the challenge of personalized e-commerce search at CIKM Cup 2016. The goal of this competition is to predict search relevance and re-rank the result items in SERP according to the personalized search, browsing and purchasing preferences. Based on a detailed analysis of the provided data, we extract three different types of features, i.e., statistic features, query-item features and session features. Different models are used on these features, including logistic regression, gradient boosted decision trees, rank svm and a novel deep match model. With the blending of multiple models, a stacking ensemble model is built to integrate the output of individual models and produce a more accurate prediction result. Based on these efforts, our solution won the champion of the competition on all the evaluation metrics.
\end{abstract}

%
%


%
%

%
%



\section{Introduction}
With the rapid development of internet and various online e-commerce services, more and more people use e-commerce search engines. However, the huge amounts of information makes it hard  to provide users the desired search results. Therefore, it is of great importance to understand the users' actual intentions and show the most preferred information to them.

At this challenge~\footnote{\small{~https://competitions.codalab.org/competitions/11161}} of CIKM Cup 2016, the DIGINETICA company~\footnote{~\small{http://diginetica.com/}} provides a dataset for academic and industry researchers to test out new ideas on the personalized e-commerce search. The dataset contains a large amount of anonymized search and browsing logs, product metadata and anonymized transaction history. The goal of this challenge is to predict search relevance of products according to users' personalized search, browsing and purchasing preferences. In the search challenge, both the ``query-full" (search engine result pages returned in response to a query) and ``query-less" (search engine result pages in response to the user click on some product category) scenarios are considered. The final evaluation performance is a weighted sum of results on ``query-full" and ``query-less" scenarios.

\begin{figure}
\centering
\includegraphics[width=0.46\textwidth]{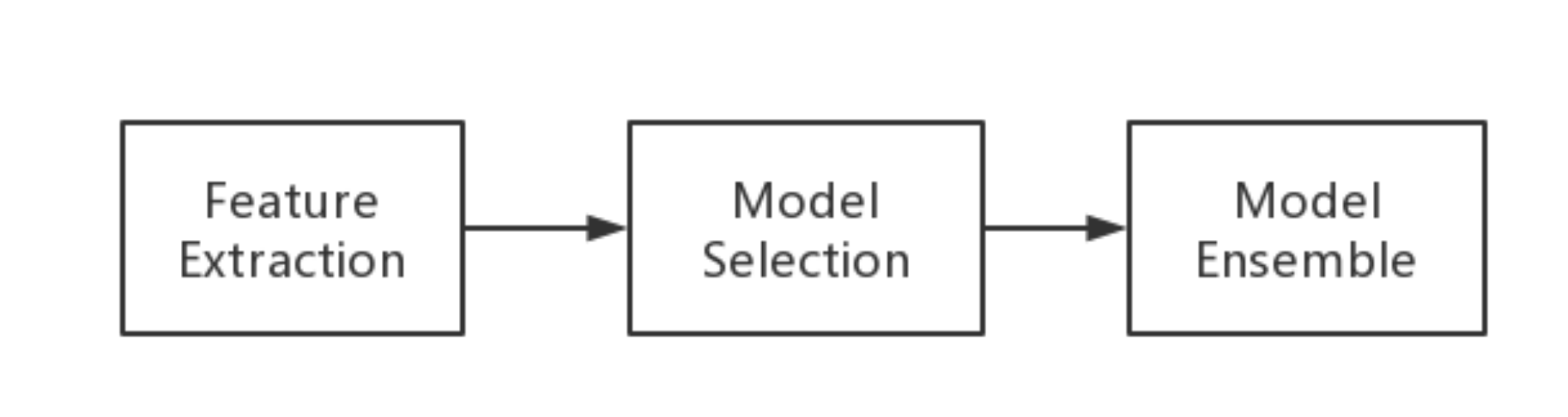}\vspace{-2mm}
\caption{the flowchart of our solution.}\vspace{-3mm}
\label{fig:flowchart}
\end{figure}

This paper describes our solution at the competition. We formulate the task as a ranking problem and adopt the learning to rank approach. A simple flowchart of our solution is shown in Fig.~\ref{fig:flowchart}. Feature engineering is one of the most critical factors for data mining problem. To better cover the key information in each query-item pair, we extracted the features from three different levels based on a preliminary analysis on the dataset, i.e., statistic features, query-item features and online session features. Then we carefully selected a series of learning to rank models and a deep match model with respect to the extracted features, which consists of our single model-based methods in the first layer. Finally, with the predicted model scores as input, we design a stacking ensemble method in the top layer to combine different models and obtain the final prediction.

The remainder of this paper is organized as follows. Section 2 briefly reviews the data set. Section 3 formulates the problem, and section 4 describes our feature engineering efforts. Section 5 presents the model framework. The results is shown in section 6, followed by the conclusion in section 7.

\section{Dataset Overview}
There are six data files in total provided for the competition. The first one is train-queries.csv, which contains all the users' search logs from Jan 1st, 2016 to June 1st, 2016. Each query log contains the corresponding search tokens (or search category for query-less scenario), session id, user id, event date, presented products, session time frame, etc. Another three files records the user's click logs, browse logs and purchase records with time stamp. The last two files contain the product metadata, such as product category, description and price. The basic statistics of the dataset are shown in Table~\ref{tab:stat1}. Table~\ref{tab:stat2} shows the user statistics of the dataset. We can see that there exist large amounts of anonymous users in our dataset (even larger than the real users with user ID). Moreover, for the real users, the user overlap between train and test queries is rather small, which brings more difficulty to the personalized search challenge.

\begin{table}[t]
\centering
\caption{\label{tab:stat1} Basic statistics of the provided dataset.}
\begin{tabular}{c|c}
\hline
    Statistics & Value \\
\hline
    \#query-full queries & 53,427 \\
    \#query-less queries & 869,700\\
    \#sessions & 573,935\\
    \#presented products & 130,987 \\
    \#click logs  & 1,127,764\\
    \#view logs & 1,235,380\\
    \#purchase records & 18,025\\
\hline
\end{tabular}
\end{table}

\begin{table}[t]
\centering
\caption{\label{tab:stat2} User statistics of the provided dataset.}
\begin{tabular}{c|c}
\hline
    Statistics & Value \\
\hline
    \#real users & 232,817 \\
    \#anonymous users & 333,097\\
    \#train real users & 140,387\\
    \#test real users & 116,630 \\
    \#(train real users $\bigcap$ test real users)  & 24,201\\
\hline
\end{tabular}
\end{table}

%
%

\section{Related Work}
We formulate this task as a learning to rank problem. We build models for each scenario (e.g., logistic regression) and utilize the ensemble of various models to further improve performance. We also introduce the deep learning model to find some implicit features. More specifically, in the query full scenario, we take the rank problem as learning to match, where we treat user query as subject and items in SERP as objects. Besides the important independent features of item and query, we also model the interaction between the subject and object. The interaction could be formed in two aspects, one is feature, .e.g., the cross features between query tokens and item tokens, and the other is model, using semantic model to represent. These two methods could been combined into the final model.

As for the query less scenario, since the query is missing, and only category is provided, it's hard to find the proper subject to form the matching approach. With some data analysis, we found it is very useful to model the short-term interests of users. In this competition the short-term interesting could be extracted from session, e.g., revisiting items~\cite{cockburn2003revisit}.

\subsection{Learning to rank}
Learning to rank has been a hot research direction in information retrieval for many years and the techniques have been widely used in the practical web search and e-commerce search applications. Depending on the generation of training samples, there are roughly three kinds of learning to rank approaches: pointwise approach, pairwise approach and listwise approach. The pointwise approach focuses on predicting the exact relevance degree of a single instance to a query, and the score of relevance degree is then used to rank the instances. Typical pointwise approaches include Pranking~\cite{crammer2001pranking}, Large Margin Ranking~\cite{shashua2002ranking} and Constraint Ordinal Regression~\cite{chu2005new}. The pairwise approach considers the relative relevance degree among different instances in terms of the same query, and use a pair of instances to act as a training sample. A large amount of research work has been proposed in this field, ranging from Ranking SVM~\cite{herbrich1999large}, RankBoost~\cite{freund2003efficient} to RankNet~\cite{burges2005learning}. The third listwise approach directly optimizes over the whole instance list under a query. Typical listwise approaches includes LambdaRank~\cite{quoc2007learning}, ListNet~\cite{cao2007learning} and LambdaMART~\cite{burges2010ranknet}. The LambdaMART model performs well in the web search ranking scenario and is the winner solution of the famous Yahoo Learning to Ranking Challenge. In our solution, we mainly adopt the first two types of learning to rank approaches.

\subsection{Ensemble learning}
The ensemble learning method has been a popular method in many real-world applications. The basic idea of ensemble learning is to combine a set of weak learners to construct a strong learner. An overview of various ensemble methods can be found in~\cite{dietterich2000ensemble}. Bagging and boosting are two most famous types of ensemble mechanisms, and some popular ensemble methods include Adaboost~\cite{ratsch2001soft}, which is used mainly for classification, and GBDT~\cite{friedman2001greedy}, which performs well in both regression and classification tasks. In our framework, we adopt LR, RankSVM, DMM as sub-learner and use GBDT as meta-learner.

Stacked generalization~\cite{wolpert1992stack} is a way of combining multiple models, which introduces the concept of a meta learner. Although an attractive idea, it is less widely used than bagging and boosting. Unlike bagging and boosting, stacking is normally used to combine models of different types.

\section{Feature Extraction}

Feature engineering is a very important task for machine learning applications. In this task, we have designed a set of comprehensive features from different levels. Given a query-item pair, we extract a set of features based on the provided search and browsing logs, query and product metadata, as well as anonymized transaction data.

\subsection{Statistic Features}

\subsubsection{global statistic features}
In the e-commerce search scenario, products' historical click/sale conditions are quite important. The first type of features are generated by counting the basic statistical information of each presented product, from the users' overall search and browsing logs, as well as the transaction data. The main statistic features are listed as below,
\begin{itemize}
  \item the total show count, click count, view count and purchase count of each product
  \item the total distinct user count of the four types of behaviors on each product
  \item the click-through rate (\emph{ctr}), view rate and click value rate (\emph{cvr}) of each product
  \item the word length of each product
  \item the percent rank of the presented product's position
\end{itemize}

Since the price feature acts poorly as a single statistic feature, we use it combined with some other absolute product statistic features, given as below:
\begin{equation}\label{equ:1}
product\_price\_feat(A) = \frac{{\# product\_behavior(A)}}{{price + 1}}
\end{equation}
where A can be click behavior, browse behavior or purchase behavior, respectively. Actually, this kind of price feature represents the cost performance ratio of the typical product.

\subsubsection{Time-based statistic features}
The popularity of products in e-commerce website changes with time, and the typical user interest also drifts. Therefore, we also devise a set of local statistic features by splitting the whole time space into multiple sessions with equal interval, and counting the product statistics in each session separately. After this, for each product statistic $v$ (e.g. click count, \emph{ctr}), we can extract a set of corresponding local statistic features $\mathbf{x}_v^{local} = \{ {x_{v,1}},{x_{v,2}},...,{x_{v,L}}\} $, where ${x_{v,l}}$ indicates the value of the local product statistic $v$ in the $l$-th time session. For the length of each time session, we make a choices of a week interval, half-month interval, month interval and two-month interval.

\subsection{Query-Item Features}

\subsubsection{Category-based token features}
Some semantic words in the product descriptions can also catch users' attention (e.g., ``promotion", ``discount"), by locating on this kind of words can also bring some benefit to the prediction performance. Besides, for different product categories, the corresponding ``click-most" words may differ. Therefore, for each product $i$, we design a kind of category-based binary token features $\mathbf{x}_{tok}^{cate} = \{ x_{{t_1}}^{{c_1}},x_{{t_2}}^{{c_1}},...,x_{{t_N}}^{{c_M}}\} $, where $x_t^c=1$, if word $t$ exists in the product description and the product category is $c$, otherwise $x_t^c=0$. Through simple feature selection mechanism, we can capture the ``click-most" words for each category.

\subsubsection{Cross token features}
The semantic relevance between the query and product is quite significant in the query-full scenario of e-commerce search, and it can be captured by measuring the co-occurrence of the query tokens and product descriptions. The typical BM25~\cite{robertson2009probabilistic} works well in the practical search engines. However, in our scenario, there exists little word overlap between query and product descriptions (about 1.8\%), which makes the BM25 not applicable. Therefore, we extract two kinds of cross-product word features between each query-product pair to measure the corresponding semantic relevance.

\begin{itemize}
  \item Word level: for the first cross-product feature, we directly generate a binary cartesian product between the single words of query and product description. Specifically, for each query and product pair $(q, p)$, we extract a binary word-word feature vector over all the word pairs between the query and product. In this way, even when the query words and product words don't overlap with each other and there exists no exact match between them, we can still predict their semantic relevance by identifying some frequently co-occurred word pairs between query and product description.
  \item Vector level: since the word overlap is small, we further adopt a word embedding method to represent the query words and product words in a shared semantic space, by leveraging a deep match model (introduced in the next subsection). In this way, each word or text description can be represented as a vector and matched in the same vector space. 
\end{itemize}

\subsection{Session features}
Repeated pattern is a common phenomenon for users' online behaviors, especially for the repeated consumption pattern that users may view/click the same product again and again during the same query session. This phenomenon is also frequently occurred in our dataset and more than 20\% clicks (in the last query) in the same session are repeated clicks. Therefore, we further design two kinds of session features to embed this typical repeated phenomenon.

\begin{itemize}
  \item For each query-product pair, we extract some binary session feature as whether or not the product has been clicked or viewed or purchased in the same session before.
  \item To better enhance the importance of the historically clicked/viewed/purchased products in each session, we further add a constant session bias to all these products in the same session when predicting.
\end{itemize}

\section{Model Framework}

\subsection{Model Selection}
We compare the performance of several models, including Gradient Boosted Decision Tree (GBDT), Logistic Regression (LR), Deep Neural Network (DNN), and RankSVM. A comparative table of all approaches are shown in result section, from which we can observe that LR with compound feature and DNN with vector features outperforms the other models. Logistic regression model is simple and much easier to tuning with different features. Deep learning could simplify feature engineering and automatically learn feature representations.

\begin{figure}
\centering
\includegraphics[width=0.46\textwidth]{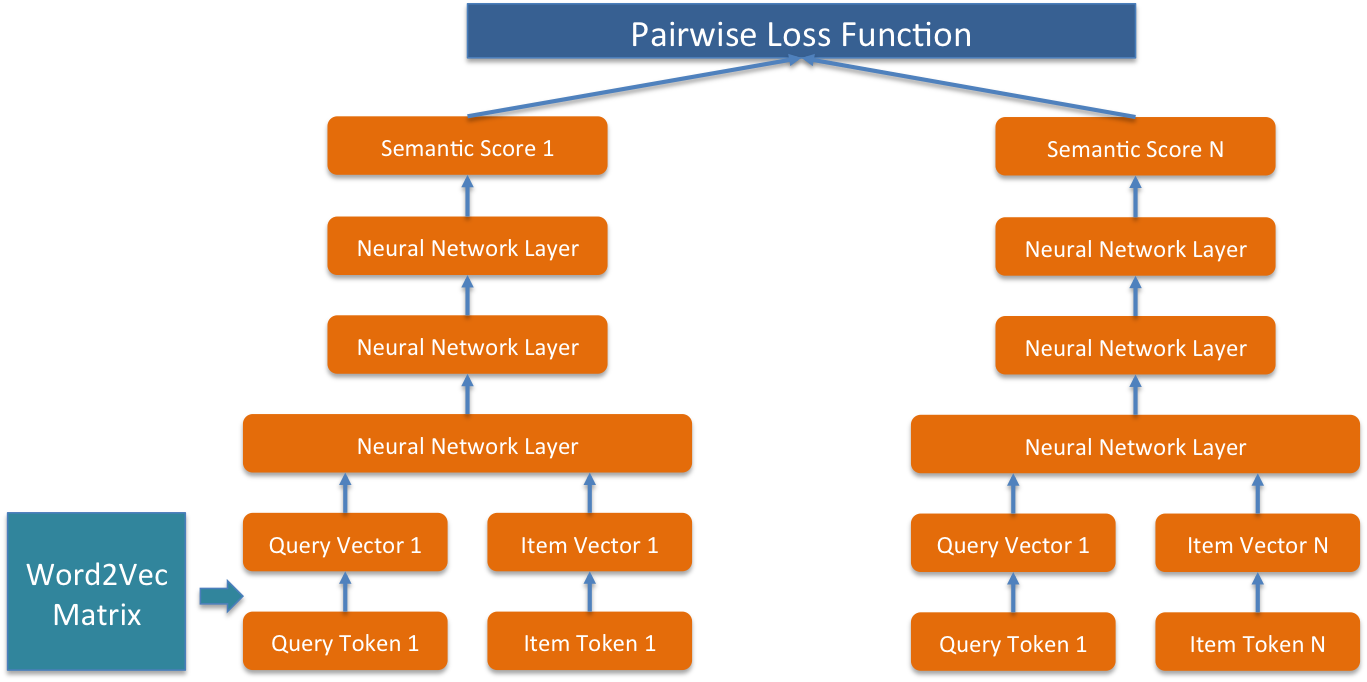}\vspace{-2mm}
\caption{the deep match model.}\vspace{-3mm}
\label{fig:dmm}
\end{figure}

In query full scenario, at first, we want to use probabilistic model e.g. BM25 to find the matching score between query and item, but with the experiments, we found the overlap between query token and item token is very limited, so we decide to search for another way to represent the query and item to make them compared. Word2vec is a good way to represent the query and item, but with unsupervised learning, the feedback information (e.g., click or pay action) is missing, so we adopted a supervised  deep match model that is shown in Fig.~\ref{fig:dmm}. With the feedback information, pairwise learning is used to train this model.

\subsection{Ensemble Modeling}
A popular way for improving performance of machine learning task is to train various models and make ensembles. However, it is usually time consuming to find an optimal way to ensemble. we use a simple but turns out to be efficient way to ensemble with stacking method, from the training data we train several models e.g. LR, RankSVM, DNN and GBDT for each scenario. 

In query full scenario, we use LR to represent relevance and DNN to represent semantic in second layer, and use GBDT as meta learner at top layer, It shows that the simple approach greatly improves the accuracy of the prediction. 

In query less scenario, the same ensemble framework is used. In addition, we do a little change in the second layer and add a extra layer, we add a model selector in second layer, the selector is used to get the model with the highest nDCG in validation set of a specific category, and logistic regression is used to train the sub-learner which used by model selector in the third layer.

Due to time limits we did not spend more effort on the model ensemble and the result implies that there are still opportunities to make improvements.

\subsubsection{Query Full}
\begin{figure}
\centering
\includegraphics[width=0.46\textwidth]{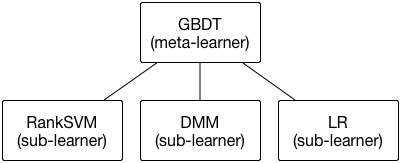}\vspace{-2mm}
\caption{the framework of query full.}\vspace{-3mm}
\label{fig:queryfull}
\end{figure}

Three models are been used in this scenario, RankSVM with aggregative feature, DNN with token feature, LR with cross query-item feature, and GBDT is used as meta learner, the framework is shown in Fig.~\ref{fig:queryfull}.

\subsubsection{Query Less}
\begin{figure}
\centering
\includegraphics[width=0.46\textwidth]{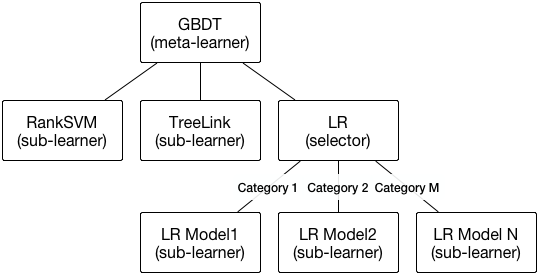}\vspace{-2mm}
\caption{the framework of query less.}\vspace{-3mm}
\label{fig:queryless}
\end{figure}

In query less scenario, we adopted RankSVM, TreeLink, lots of LR models, and a LR Model selector, the framework is shown in Fig.~\ref{fig:queryless}. Since each SERP is associated with a specific category, a raw concept is to build model for each category. 

With data analyzing, we found the data distribution is vary in category, some categories may not have enough data for training, so we change to another way, a bagging-like method, we build several LR models with same feature slot but different combination methods. The goal of this method is to promote NGCG, so we calculate category NDCG for each LR model in the test set, then use the simple strategy that the model with max NDCG will be selected as the final model. When a search query been received, we get the category from the query and use model selector to select the model with max category NDCG among the candidates.

\section{Results}
The popular ranking metric NDCG is used to evaluate the final results. The query-full and query-less queries are evaluated separately. The final NDCG score is a weighted sum of the query-full NDCG and query-less NDCG as:
\begin{equation*}
  NDCG = 0.8*NDC{G_l} + 0.2*NDC{G_f}
\end{equation*}
where $NDC{G_l}$ and $NDC{G_f}$ are the NDCG score for query-less and query-full queries, respectively.

Table~\ref{tab:1} shows the results of different models in our solution on the validation board. Note that due to the storage limit, the token features are not used in our Ranking SVM and GBDT models. We can see that: (1) by only leveraging query words and product description words, our deep match model (DMM) can obtain a relatively good NDCG score (0.5077) on the query-full set, which shows the effectiveness of this method on capturing the correlation between the semantic words of query and product~\footnote{\small{The performance of DMM is inferior to that of LR, may be due to the limited data volume in query-full scenario.}}. (2) the LR model performs the best among all the single model-based methods, demonstrating its ability to combine many different kinds of features, e.g. one-level statistic features and two-level cross-product features. (3) the best result is obtained with the ensemble methods, which may be due to that different models can complement each other in the final decision.

\begin{table}[t]
\centering
\caption{\label{tab:1} Validation results of different models in our solution.}
\begin{tabular}{ccccc}
\hline
    Ranking Model & NDCG & NDCG full & NDCG less \\
\hline
    Ranking SVM & 0.3633 & 0.4197 & 0.3492\\
\hline
    GBDT & 0.3650 & 0.4291 & 0.3489\\
\hline
    DMM & - & 0.5077 & - \\
\hline
    LR (all feats)  & 0.4175 & 0.5473 & 0.3851\\
\hline
    Ensemble & 0.4238 & 0.5548 & 0.3911\\
\hline
\end{tabular}
\end{table}

\section{Conclusions}
In this paper, we describe our solution for the CIKM Cup 2016 Personalized Search Challenge at which we took the first place. The solution has three primary components: data construction, feature engineering, and ensemble modeling. Due to limited time, there are still some possible methods that have not been fully explored. e.g., recommender system is also a promising approach to this competition, and other ensemble methods could also be worth trying.

\section{Acknowledgments}
Thank the organizers of CIKM Cup 2016 and DIGINETICA for generously providing the opportunity and data resources for training and testing our solutions. 

%
\bibliographystyle{abbrv}
\bibliography{sigproc}  
%
%

\end{document}